\documentclass{article} % For LaTeX2e
\usepackage{arxiv,times}

\usepackage{amsmath,amsfonts,bm}

\def\eqref#1{equation~\ref{#1}}
\def\1{\bm{1}}

\DeclareMathAlphabet{\mathsfit}{\encodingdefault}{\sfdefault}{m}{sl}
\SetMathAlphabet{\mathsfit}{bold}{\encodingdefault}{\sfdefault}{bx}{n}

\usepackage{hyperref}
\usepackage{url}
\usepackage{pifont}
\newcommand{\cmark}{\textcolor{green!60!black}{\ding{51}}}
\newcommand{\xmark}{\textcolor{red!70!black}{\ding{55}}}
\newcommand{\pmark}{\textcolor{orange!90!black}{$\sim$}}

\title{PANDA: A Decentralized Architecture with Flexible Orchestration for Scalable, Fault-Tolerant Multi-Agent Systems}

\author{{Matthew D. Laws, Cristina Nita-Rotaru} \\
Khoury College of Computer Sciences\\
Northeastern University\\
\texttt{\{laws.ma,c.nitarotaru\}@northeastern.edu} \\
}

\usepackage{xspace}
\usepackage{graphicx}

\newcommand{\panda}{\textsc{\textbf{panda}}\xspace}
\newcommand{\pandastar}{\textsc{\textbf{panda-star}}\xspace}
\newcommand{\pandachain}{\textsc{\textbf{panda-chain}}\xspace}
\newcommand{\pandamesh}{\textsc{\textbf{panda-mesh}}\xspace}

\usepackage{booktabs}
\usepackage{subcaption}

\iclrfinalcopy % Uncomment for camera-ready version, but NOT for submission.
\begin{document}

\maketitle

\begin{abstract}
Existing architectures for LLM-based multi-agent systems (MAS)
cannot reliably and efficiently solve multi-step tasks at scale: they struggle to support large numbers of agents and concurrent tasks, tolerate failures, govern agent interactions, and accommodate the diverse planning and execution patterns different tasks require.
We present \panda, a decentralized architecture that connects a large collective of heterogeneous, independently administered agents, letting them discover each other's capabilities and self-organize into small specialized teams per task. \panda scales by decoupling collective communication from team communication, allowing agents to participate in multiple teams simultaneously, load-balancing tasks across the collective, and scheduling concurrent work within each agent. 
\panda further separates the underlying architecture from the orchestration strategy, supporting three planning and execution patterns (star, chain, and mesh) that can be selected according to the structure and requirements of each task. \panda detects infrastructure and orchestration failures and recovers affected tasks by dynamically replanning around failed components. Finally, to provide governance without a centralized service that would limit scalability, \panda uses a web-of-trust model to constrain agent interactions to established trust relationships.
We evaluate \panda on the HotPotQA benchmark, demonstrating that it scales to thousands of agents, assembles teams in milliseconds, matches state-of-the-art accuracy at up to $8\times$ the efficiency, and sustains 100\% task completion under faults where existing systems fail.
\end{abstract}

\section{Introduction}

LLM-based agents are increasingly specialized through tools, knowledge, and skills for particular tasks. However, real-world problems often involve multiple steps spanning different domains and therefore require capabilities beyond those of any single specialized agent. Multi-agent systems (MAS) enable agents with complementary capabilities to divide work, exchange intermediate results, and collaborate, making them a promising paradigm for solving complex, multi-step tasks.

Realizing this potential requires MAS to support large, dynamic collectives in which many agents may concurrently collaborate on different tasks. We identify four core requirements for such systems: (1) scalability, (2) fault tolerance, (3) flexible orchestration, and (4) governance. Scalability requires supporting growth in the number of agents, concurrent tasks, and interactions while maintaining task utility and controlling the communication and token costs incurred to achieve it. Fault tolerance requires tasks to continue despite failures of agents or the underlying infrastructure. Flexible orchestration requires supporting multiple planning and execution models to accommodate different tasks and teams. Lastly, governance requires establishing trust and constraining interactions to trusted agents as collectives extend beyond a single administration. 

MAS are difficult to scale because centralized agent registries, orchestration, and governance providers become bottlenecks as the number of agents and tasks grows. At the same time, agent discovery, coordination, communication, and shared-state management generate increasing latency and token costs. At scale, component failures become more likely, further increasing recovery overhead, causing system costs to grow faster than task utility.

\textbf{Limitations of previous work.} 
First, current MAS architectures have limited scalability.
Centralized designs such as Magentic-One \citep{magentic-one} concentrate planning and coordination in a single component, creating a bottleneck as the number of agents and tasks increases. Decentralized alternatives also face scalability limitations: AgentNet \citep{AgentNet} requires every agent to initially know all other agents and then prunes connections over time, and its decentralized planner incur high agent-selection costs when choosing among thousands of candidates at each step. Symphony \cite{symphony} and Internet of Agents \citep{IoA} are also limited by centralized communications hubs.

Second, existing governance solutions rely on centralized authorities, limiting their scalability and applicability to decentralized deployments. SAGA \citep{saga_ndss2026} and MAGIQ \citep{MAGIQ} rely on a provider to establish agent identities and enforce policies, and \citet{saga_shield} extends the provider to be resilient in a distributed trust setting, but the architecture remains logically centralized. As the number of agents and interactions grows, the authority becomes a bottleneck and a single point of failure. Moreover, such an authority may not exist in decentralized settings, where no single party can be trusted to authenticate, authorize, or vouch for all participating agents.

Third, existing MAS architectures provide limited orchestration flexibility. Magentic-One and MetaGPT \citep{metagpt}, impose centralized planning mechanisms, limiting their ability to support different planning and execution models across tasks. Internet of Agents \citep{IoA} supports more flexible coordination, but routes interactions through a centralized server, limiting its scalability. MACNET \citep{MACNET} supports different orchestration patterns but requires agents to be manually organized before execution, preventing runtime flexibility.

Finally, existing MAS provide only limited fault tolerance. Prior work handles model and communication errors through retries, fallback responses, redundancy, or consensus \citep{AgentScope_2024, AgentNet, mas_bft_2026}. However, these approaches do not provide comprehensive recovery from infrastructure and orchestration failures, including preserving execution state, reassigning unfinished work, and continuing partially completed workflows. Consequently, failures can still disrupt execution or require tasks to restart at higher token and communication cost.

\textbf{Our contribution.} We present \panda (\textbf{P}lanning \textbf{A}gents in a \textbf{N}etworked \textbf{D}ecentralized \textbf{A}rchitecture), a decentralized architecture that connects a large collective of heterogeneous, independently administered agents, letting them discover each other's capabilities and self-organize into small specialized teams per task. \panda scales by decoupling collective communication from team communication, allowing agents to participate in multiple teams simultaneously, load-balancing tasks across the collective, and scheduling concurrent work within each agent. 
\panda further separates the underlying architecture from the orchestration strategy, supporting three planning and execution patterns (star, chain, and mesh) that can be selected according to the structure and requirements of each task. \panda detects infrastructure and orchestration failures and recovers affected tasks by dynamically replanning around failed components. Finally, to provide governance without a centralized service that would limit scalability, \panda uses a web-of-trust model to constrain agent interactions to established trust relationships. Although decentralized trust and reputation models have been studied in traditional MAS, we are unaware of any LLM-based MAS architecture that uses a web of trust for decentralized governance.
We summarize the contributions of \panda over previous work in Table~\ref{tab:contributions}.

\begin{table}
\centering
\caption{Comparison of \panda with existing multi-agent systems. \cmark~= supported, \pmark~= partially supported, \xmark~= not supported. Flexible refers to flexible orchestration patterns.}
\label{tab:contributions}

\setlength{\tabcolsep}{5pt}
\renewcommand{\arraystretch}{1.25}
\small
\begin{tabular}{@{}lccccc@{}}
\toprule
System & Scalable & Fault Tolerant & Flexible & Governance & Decentralized \\
\midrule
Magentic-One \citep{magentic-one} & \xmark & \xmark & \xmark & \xmark & \xmark\\
Internet of Agents \citep{IoA} & \pmark & \xmark & \pmark & \xmark &\xmark \\
Symphony \citep{symphony} & \pmark & \pmark & \xmark & \xmark & \pmark \\
MACNET \citep{MACNET} & \pmark & \xmark & \pmark & \xmark & \cmark\\
AgentNet \citep{AgentNet} & \pmark & \pmark & \xmark & \xmark &\cmark\\
\midrule

\panda (our solution) & \cmark & \cmark & \cmark & \cmark & \cmark \\
\bottomrule
\end{tabular}
\end{table}

\textbf{Results.} We evaluate \panda on the HotPotQA benchmark \citep{hotpot} and demonstrate three key properties across all topologies. (1) \panda scales efficiently to large collectives under heavy workloads, assembling teams in milliseconds with per-assembly cost that remains constant even at $10^6$ agents, and solving tasks up to $8\times$ faster than baselines while achieving comparable accuracy. (2) \panda is resilient to both infrastructure and orchestration faults, maintaining high accuracy and 100\% task completion where existing systems fail. (3) \panda's decentralized governance establishes tunable trust across the collective, letting agents configure how far trust propagates while maintaining $>99\%$ precision under churn.

\section{PANDA Architecture}
\label{sec:arch}
In this section we present the architecture of our system, \panda. We first present a high-level overview and then provide details about the main components.

\begin{figure}
    \centering
    \includegraphics[width=0.7\linewidth]{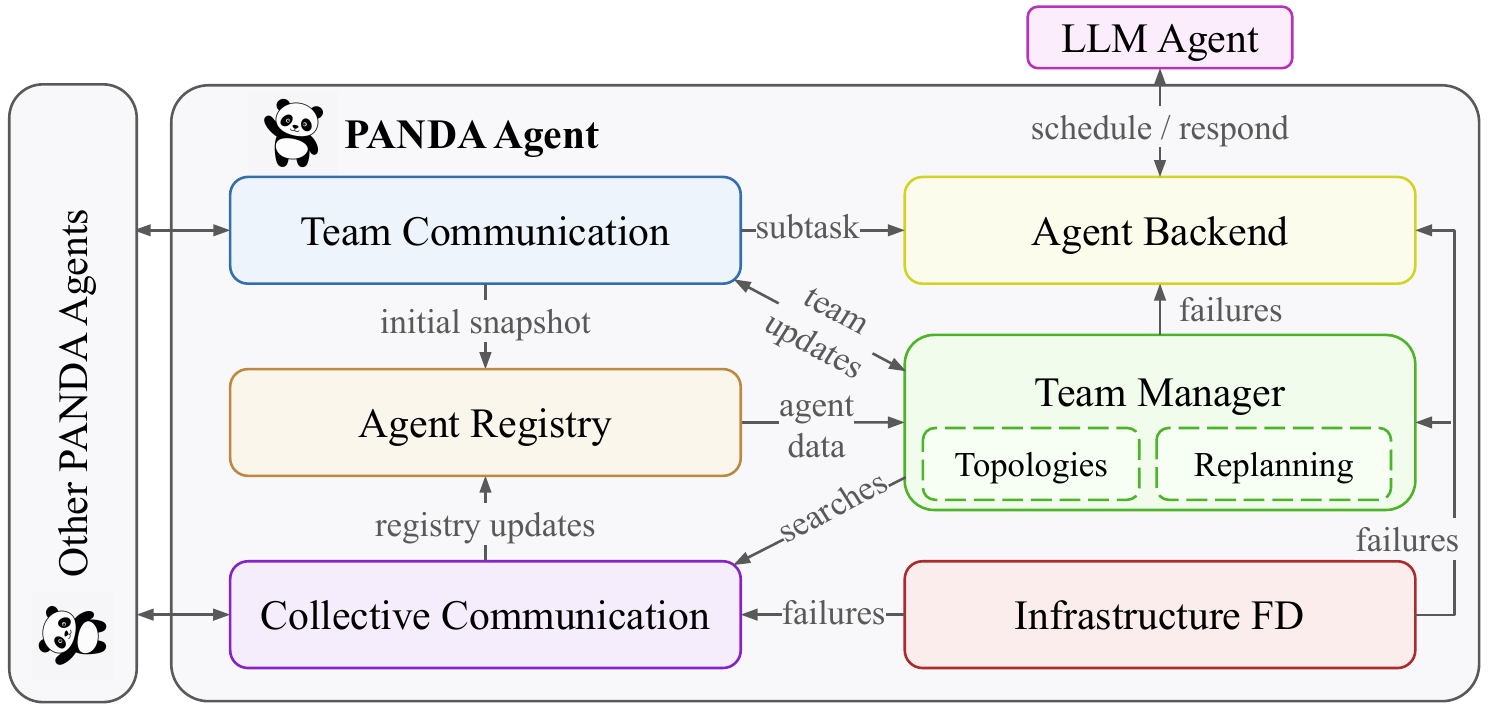}
    \caption{Internal design of a \panda agent. \emph{Team Communication} handles incoming and outgoing messages for any teams the agent is a member of. The \emph{Agent Registry} maintains state about other agents in the collective, including their capabilities and governance information. \emph{Collective Communication} manages propagation and reception of system-wide messages such as agent joins and leaves. The \emph{Team Manager} assembles teams and maintains state about currently active ones. It also possesses two subunits that handle the logic for the different planning and execution topologies and replanning when failures occur. The \emph{Infrastructure Failure Detector (FD)} detects failures and relays them both internally and to the rest of the collective. The \emph{Agent Backend} is our wrapper around any custom or existing \emph{LLM Agent} that maintains stateful information relevant to \panda about the agent. In turn, the agent must expose a list of capabilities $C$ and implement a function $f(c \in C, \mathrm{request}) \mapsto \mathrm{respose}$.}
    \label{fig:panda_node}
\end{figure}

\subsection{Design Goals and Overview}

\panda is a MAS designed around four central goals. 
\textbf{(1)~Scalability:} \panda must handle massive numbers of agents and concurrent tasks, with per-task cost that grows sub-linearly with respect to the collective so that adding agents expands capability without degrading throughput. 
\textbf{(2) Governance:} Agents should be able to establish trusted connections and control who they interact with within the collective. 
\textbf{(3)~Flexible planning and execution patterns:} Rather than committing to a single topology, \panda must support heterogeneous agents and arbitrary communication patterns so that each task can be solved with the structure best suited to it.
\textbf{(4)~Fault tolerance:} Agents may crash, become unreachable, return malformed outputs, or infrastructure links may drop; \panda must detect these conditions and complete tasks reliably despite them, without operator intervention.  

To provide scalability and eliminate single points of failure,
\panda adopts a decentralized (peer-to-peer) architecture. \panda provides the infrastructure and abstractions that allow independently created agents to discover other agents, establish teams, and coordinate to solve complex tasks. Figure~\ref{fig:panda_node} visualizes the internals of a single \panda agent. \panda supports any LLM agent, as long as it implements the function required by Agent Backend module. This module manages \panda-specific state for the agent and handles local subtask scheduling, since an agent can be part of multiple teams working on separate tasks concurrently. Communication takes place over two separate services, a collective communication service, optimized for scalability, and a team communication service, optimized for reliability. More details about scalability mechanisms are presented in Section \ref{sec:scalability}.

There is no central authority or central repository in \panda. Instead, each \panda agent maintains a local registry tracking a subset of other agents, their capabilities, how to reach them, and the information required for governance (public key certificate). The registry is initialized when an agent joins the collective through a current member, and updated as the collective state changes. To enable sublinear capability lookups, each registry maintains an inverted index mapping capabilities to the agents that provide them. Agents establish trust using a self-governance mechanism based on web-of-trust and described in detail in Section \ref{sec:governance}. 

\panda decouples the communication topology from the orchestration logic, to ensure efficient task planning and execution. It supports three (star, chain, and mesh) topologies and the \emph{Team Manager} allows the selection of the appropriate orchestration strategy  handling the logic for planning and execution. Details about the different orchestration mechanisms are presented in Section~\ref{sec:orchestration}.

\panda provides support for infrastructure and orchestration failure detection and recovery. Each agent contains an Infrastructure Failure Detector (FD) module that detects failures and relays them both internally and to the rest of the collective, along with mechanisms for swapping out crashed or misconfigured agents. Failure detection and replanning strategies are described in Section~\ref{sec:failure_detection}.

\subsection{Scalability}
\label{sec:scalability}
\panda targets scalability through multiple mechanisms: (1)  decentralized registries and governance; (2) highly-efficient communication, scheduling, and load balancing services; (3) flexible orchestration beyond centralized approaches; (4) fault-tolerant team assembly, task planning, and execution. Below we present the system level scalability mechanisms, and describe the governance, flexible orchestration, and fault tolerance mechanisms in the following subsections.

\textbf{Communication services.}
Maintaining a full mesh of reliable connections between agents is intractable at scale. \panda decouples communication into two services, a collective communication service optimized for scalability and relying on gossip protocols (our implementation uses GossipSub) \citep{gossipsub} and a team communication service that is instantiated on the fly for each team and is designed for reliability (our implementation uses TCP).

\textbf{Scheduling.}
To minimize idle time, \panda agents can belong to multiple teams simultaneously. To facilitate this, we introduce a scheduling mechanism inspired by a classical CPU. Instead of immediately executing assigned subtasks, agents enqueue their work, and an internal scheduling policy dispatches subtasks as the underlying agent becomes available. We support first-in-first-out (FIFO) and lottery scheduling, but more sophisticated policies can be added. 
This is distinct from LLM serving \citep{Autellix, fast_serve} as it operates at the agent level.

\textbf{Load Balancing.}
To prevent any single agent from being overloaded, we employ a load-balancing mechanism adapted from \citet{sparrow}. During team assembly, for each capability $c$ with redundancy $r(c)$, we poll $d\cdot r(c)$ agents with capability $c$ and ask each how many teams they currently belongs to; the $r(c)$ least loaded of the group joins the new team. \citet{sparrow} demonstrates that a probe ratio $d = 2$ performs best under high cluster load with over $2.5\times$ faster median response time compared to random sampling ($d=1$). When $d=2$ and $r(c)=1$, this reduces to \citet{power_of_two}'s Power of Two Choices.

\subsection{Web-of-Trust Governance}
\label{sec:governance}

\panda's governance is inspired by the web of trust (WoT) introduced by PGP \citep{pgp}. WoT has been used in settings such as gating Debian Developer status \citep{debian}, but to our knowledge \panda is the first to apply it as a concrete governance mechanism for LLM-based MAS.

We formalize our governance as follows. Each agent $A_i$ holds a keypair $(pk_i, sk_i)$ from a signature scheme $(\mathsf{KeyGen}, \mathsf{Sign}, \mathsf{Verify})$. When an agent $A_i$ wants to join the collective, they must be certified by one or more existing member of the collective after providing out of band proof of their identity.\footnote{For instance, via a dedicated key-signing party or an existing real-world connection.} When an existing member $A_{j}$ certifies a new agent $A_i$, $A_{j}$ issues a certificate ($\mathrm{cert}_{j \to i}$) binding $A_i$'s key and identity together with its advertised capability set ${c}_i$ and issuance timestamp $ts$:
\begin{equation}
    \mathrm{cert}_{j \to i} = \bigl(pk_i,\; \mathrm{id}_i,\; c_i,\; ts,\; \sigma_{j \to i}\bigr), \qquad
    \sigma_{j \to i} = \mathsf{Sign}_{sk_j}\!\bigl(pk_i \,\|\, \mathrm{id}_i \,\|\, c_i \,\|\, ts\bigr).
\end{equation}
$A_i$ will also often certify $A_j$ assuming the out of band proof was exchanged mutually. 
Certificates are gossiped alongside capability announcements, so each agent's local registry contains a partial view of the certification graph, with a directed edge $A_j \to A_i$ for every known certificate $\mathrm{cert}_{j \to i}$. The collective is bootstrapped by one or more trusted \emph{genesis} nodes: a single genesis node self-certifies, or multiple genesis nodes mutually certify one another, establishing the initial trust base.

Trust is anchored at a specific agent and extended by depth-bounded reachability over the certification graph. $A_i$'s trusted set contains every peer reachable from its anchor via a chain of at most $n$ certifications. \panda supports two anchor modes. Under the \emph{self} anchor, agent $A_i$ anchors on itself. Under the \emph{genesis} anchor, $A_i$ uses a genesis node as its anchor. 
In both cases we define the trusted set for agent $A_i$, denoted $\mathcal{T}_i$, recursively as:

\begin{equation}
    \mathcal{T}_i^{(0)} = \{A_{i}^{\mathrm{anchor}}\}, \qquad
    \mathcal{T}_i^{(k)} = \mathcal{T}_i^{(k-1)} \cup \bigl\{A_x : \exists\, A_y \in \mathcal{T}_i^{(k-1)} \text{ with valid } \mathrm{cert}_{y \to x}\bigr\},
\end{equation}
and let $\mathcal{T}_i = \mathcal{T}_i^{(n)}$ for a configurable depth $n \in \mathbb{N} \cup \{\infty\}$. Each agent chooses its anchor mode and depth locally, tailoring trust to its own requirements.

\subsection{Flexible Orchestration}
\label{sec:orchestration}

\textbf{Team assembly.}
Solving multi-agent tasks efficiently requires quickly determining which agents will contribute to a task -- a process we call \emph{team assembly}. A task is submitted to an agent designated the \emph{entrypoint}, which assembles a team based on a set of required capabilities $\mathcal{C}$. This set is either provided by the user alongside the task or initialized dynamically by the entrypoint upon receiving the request, consulting its local registry to determine which capabilities are available.  
Once a team is set, an overlay network is initialized over the members corresponding to the specified communication pattern, allowing for fast, reliable communication between the team members. This approach is distinct from DyLAN \citep{dylan}, which relies on dense inter-agent communication that is intractable at scale. We formalize the assembly protocol in Appendix~\ref{app:asm_protocol}.

\textbf{Planning and execution. }\panda supports multiple execution topologies within a team, and we implement three that span the design space of state-of-the-art MAS: star, chain, and mesh. Each is equipped with a planning strategy that produces a plan $\mathcal{P}$ as a directed acyclic graph (DAG) of subtasks \citep{GoT}, where each vertex is bound to a required capability and an agent drawn from the assembled team. In the \textit{star} approach, an orchestrator creates the plan and dispatches dependency-free subtasks in parallel, synthesizing intermediate results as subtasks complete. 
In the \textit{chain} approach, agents form a sequential chain in which each communicates only with its predecessor and successor. A random initiator produces a linear plan and hands it to the first agent, each subsequent agent executes its subtask and forwards the plan onward. In the \textit{mesh} approach, a subset of \textit{collaborators} covering every capability jointly agrees on $\mathcal{P}$ via a propose--rank--revise loop: each collaborator proposes a plan, all score every proposal via ranked-choice voting, the lowest-scoring proposals are pruned, and surviving authors revise; the loop terminates when a single plan remains \citep{robustMAS}. Once a consensus plan has been reached, subtasks with redundant capabilities produce their output via multi-agent debate \citep{multi_agent_debate} and completed outputs are broadcast to the team. 
A supermajority vote after each subtask can trigger replanning under the same propose--rank--revise loop.
We depict each topology in Figure~\ref{fig:topologies} and present more details in Appendix~\ref{app:topos}.

\begin{figure}[h]
    \centering
    \includegraphics[width=0.7\linewidth]{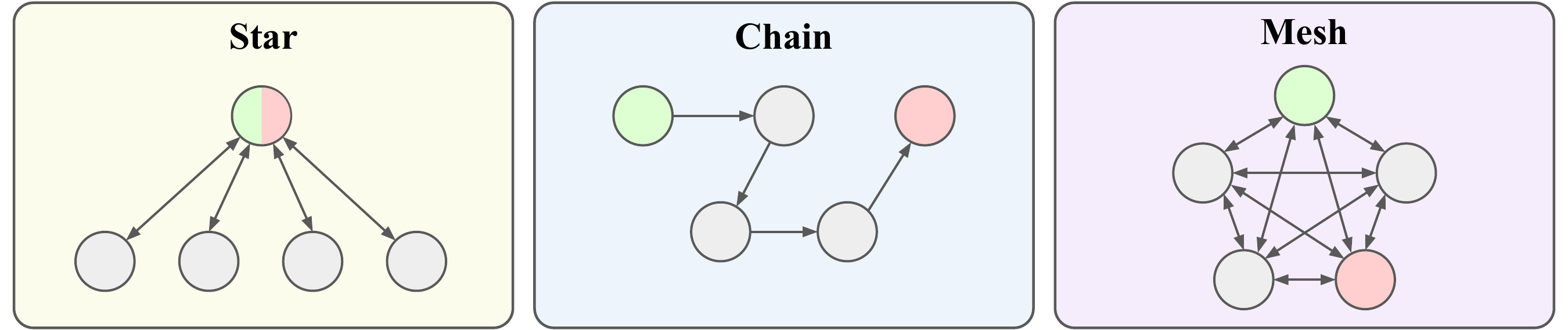}
    \caption{An example of each topology. Green indicates where input is passed to the team, and red denotes who provides the output. In the star, I/O flows through the orchestrator. In the chain, input enters the first agent and exits the last. In the mesh, I/O can begin and end at any teammate.}
    \label{fig:topologies}
\end{figure}

\subsection{Fault Tolerant Orchestration} 
\label{sec:failure_detection}
\panda provides two types of mechanisms for dealing with faults across the lifecycle of a task: \textit{proactive mechanisms} in the form of provisioning backup agents and redundant execution, and \textit{reactive mechanisms} in the form of failure detection and replanning. \panda has dedicated detection mechanisms for both infrastructure and orchestration failures. Infrastructure failures cover faults external to the agents, such as crashes and network partitions. Orchestration failures cover faults in how the agents themselves plan, organize, and execute.

\textbf{Proactive mechanisms: Agent provisioning.}
\panda provisions for redundancy to mitigate failures. The team assembly protocol accommodates
fault-tolerance by proactively recruiting backup agents for each capability needed to solve the task. Each capability $c \in \mathcal{C}$ can be annotated with a redundancy $r(c) \in \mathbb{Z}^{+}$ that denotes how many agents with capability $c$ to recruit. 

\textbf{Proactive mechanisms: Redundant execution.}
All orchestration strategies support a redundancy extension: a subtask $v$ is dispatched to $n \leq r(\mathrm{cap}(v))$ agents in parallel, and the first successful response is accepted and the rest are canceled, tolerating up to $n - 1$ failures without needing to recruit a new agent.%invoking~\texttt{swap}.

\textbf{Reactive mechanisms: Infrastructure-level failures.} 
\label{sec:}
To detect infrastructure-level failures, we adopt the scalable failure-detector model of \citet{scalable_failure_detectors}. Each agent periodically pings another and if no response arrives, it asks several other agents to ping the suspected agent and relay the results. If all such pings fail, the agent is declared dead after a timeout and an eviction message is sent to the collective using the collective communication service.  False positives are possible; however, if an agent learns of its own removal, it simply rebroadcasts itself, canceling the eviction. The expected time to detect ($\Delta_{\text{det}}$) a failure is probabilistic, given by $    \Delta_{\text{det}} = T \cdot \frac{e^{q}}{e^{q}-1}$ where $T$ is the protocol period and $q$ is one minus the per-channel message loss probability. Because we probe over TCP, delivery is reliable ($q = 1$), and the expression simplifies to $T \cdot \frac{e}{e-1}$. We can tune $T$ to balance detection speed and network traffic. 

Agents are also proactive about confirming they are still considered alive. If an agent receives no messages within a specified timeout, it pings $k$ peers to ask whether they still consider it alive. If they respond negatively, the agent rebroadcasts its capabilities to the collective; if it was a false alarm, the agent resets its timeout and continues.

\textbf{Reactive mechanisms: Orchestration failures and replanning.} 
Structuring plans as DAGs of typed subtasks enables efficient recovery by replacing the agent assigned to a failed vertex rather than restarting the entire task. When an agent fails, the system invokes the swap protocol that assigns the affected subtask to another compatible agent already in the team, when redundancy permits, or recruits a new agent otherwise. We formalize the swap protocol in Appendix~\ref{app:swap_proto}.

For the star orchestration, worker failures are detected and handled by the orchestrator. After each subtask, the orchestrator may revise the plan -- holding completed subtasks fixed -- to account for any failures encountered, invoking the swap protocol if necessary. Orchestrator failure can be monitored by the entrypoint, and improved by checkpointing the plan and its execution state to a peer for warm restarts if an orchestrator crashes.

For the chain orchestration, once agent $A_i$ forwards to its successor $A_{i+1}$, it sends \texttt{DONE} to its predecessor $A_{i-1}$; if $A_{i-1}$ detects $A_i$ failure before receiving \texttt{DONE}, it invokes the swap protocol and reissues the task to $A_i'$. Reissues carry a monotonic generation number, so duplicate chains caused by agents that continues after being swapped are killed when detected. Replanning is also local: after an agent receives a subtask, it may rewrite the remaining plan before executing.

For the mesh orchestration, planning-time failures drop the affected collaborator from contention or trigger a swap; execution-time failures either drop the failed agent from an ongoing debate or trigger  swap when the capability has no redundancy.
\section{Experiments}

We evaluate \panda across three properties: its ability to scale to large collectives under heavy workloads (Section~\ref{sec:eval_scale}), its resilience to infrastructure and orchestration failures (Section~\ref{sec:eval_fail}), and the effectiveness of its decentralized governance in establishing trust across the collective (Section~\ref{sec:eval_gov}).

\textbf{Baselines.} 
We compare \panda against Magentic-One, which uses a star topology; AgentNet, which uses a chain; and Internet of Agents (IoA), which uses a recursive-star (tree) topology. 

\textbf{Benchmark.} We evaluate on the HotPotQA benchmark in the fullwiki setting, a Wikipedia-based question-answering dataset in which each question requires reasoning across two pages. Questions take one of two forms: \emph{bridge}, where information from one page must be gathered before the final answer can be located on a second page, and \emph{comparison}, where information from two pages must be contrasted. To ensure that agents work from the dataset's original snapshot rather than more recent content, they retrieve pages through an MCP server that hosts the Wikipedia version used to curate the benchmark. Agents are explicitly instructed to answer only from evidence returned by the MCP server, and empirically when the server is offline they correctly fail rather than hallucinate. 

\textbf{Configuration.} Agents use \texttt{GPT-5.4-mini} as the underlying LLM. Each system runs with 10 worker agents under two configurations: \emph{complete knowledge}, where all agents access the full Wikipedia snapshot, and \emph{partial knowledge}, where they are split evenly across two disjoint halves.

\textbf{Metrics.} We report four metrics: exact match (EM), F1, completion, and runtime. EM and F1 follow \citet{hotpot}: EM is a post-normalization exact match between the expected and provided answers, and F1 is a per-word F1 giving partial credit for partially correct answers. Completion measures whether any answer was produced, regardless of correctness. We report runtime (RT) in several forms, all in wall-clock seconds: \emph{total} is the time for all tasks to complete, while \emph{completed}, \emph{failed}, \emph{correct}, and \emph{incorrect} are the average per-task times over tasks in each respective category.

\subsection{PANDA Scalability}
\label{sec:eval_scale}

\begin{figure}[h]
    \centering
    \begin{subfigure}[t]{0.32\linewidth}
        \centering
        \includegraphics[width=\linewidth]{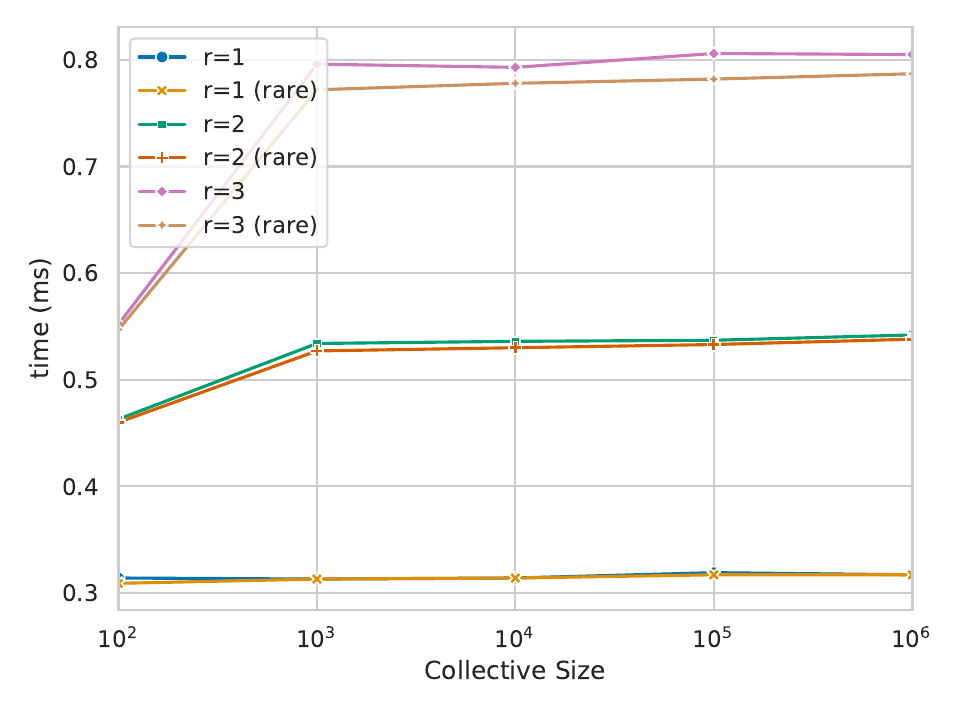}
        \caption{$c = 5$}
        \label{fig:asm_c5}
    \end{subfigure}
    \hfill
    \begin{subfigure}[t]{0.32\linewidth}
        \centering
        \includegraphics[width=\linewidth]{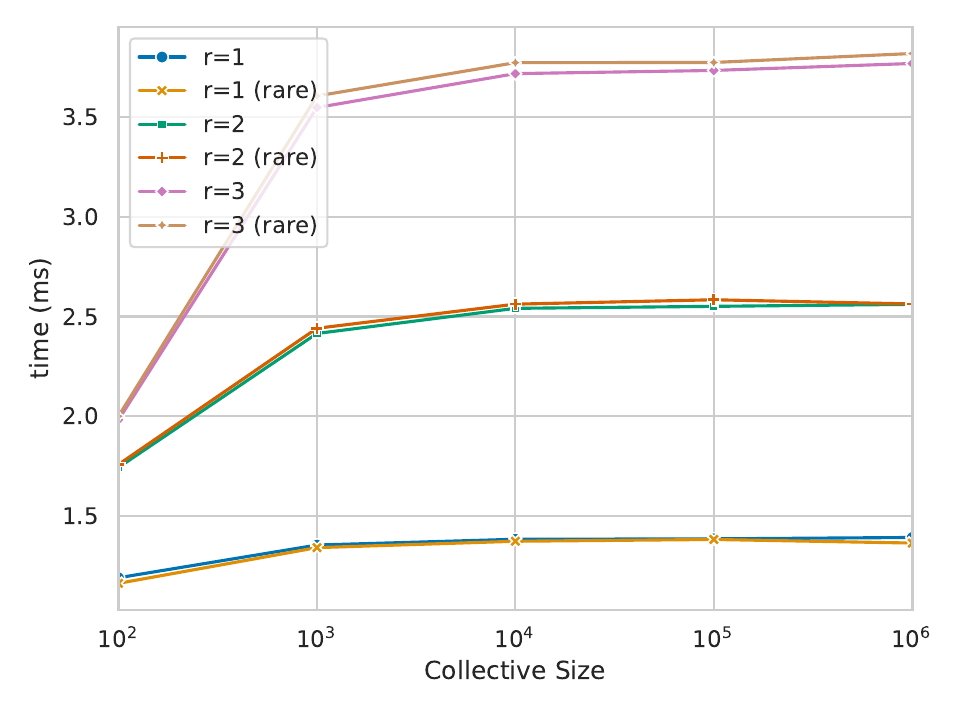}
        \caption{$c = 25$}
        \label{fig:asm_c25}
    \end{subfigure}
    \hfill
    \begin{subfigure}[t]{0.32\linewidth}
        \centering
        \includegraphics[width=\linewidth]{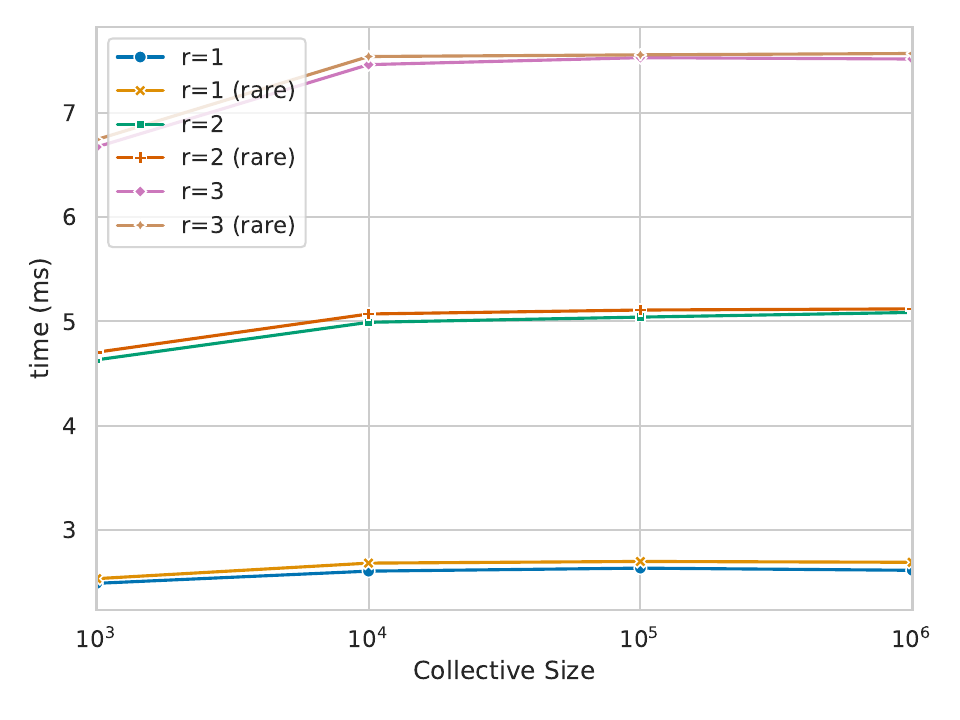}
        \caption{$c = 50$}
        \label{fig:asm_c50}
    \end{subfigure}
    \caption{Team assembly times for number of capabilities $c = 5$, $25$, and $50$ across redundancies $r =1$, $2$, $3$. Rare refers to when only 5 agents with a given needed capability exist. At $c=50$, 100 agents are not sufficient for staffing the needed capabilities.}
    \label{fig:asm_all}
\end{figure}

\paragraph{Team Assembly.}
To show that \panda generalizes to thousands of agents, we measure the cost of team assembly as a function of collective size. Once assembly completes and the team overlay is established, subsequent work is confined to the team and does not scale with the collective. We construct a collective of $n$ fully connected peers, each advertising $5$ capabilities sampled uniformly from a pool of $100$. Figure~\ref{fig:asm_all} reports median time to complete a team assembly request across $c = 5, 25, 50$ requested capabilities with redundancies $r=1,2,3$. For each configuration we include a \emph{rare} variant in which one of the required capability has only five providers in the collective.

We observe several key patterns. First, past a critical collective size -- at which sufficient redundancy across all capabilities is reached -- assembly time is constant with respect to the collective size. The initial rise reflects a smaller candidate pool, which lets assembly commit quickly; however, once coverage is sufficient, our load balancer probes a capped set of candidates regardless of $n$, giving constant scaling thereafter. The \emph{rare} variant is slightly faster for the same reason: scarce capabilities constrain the candidate set earlier. Assembly time does scale with team size, but remains on the order of milliseconds, negligible relative to the runtimes in Table~\ref{tab:hotpotqa}. Finally, team assembly incurs a fixed network overhead equal to the maximum RTT between the assembling agent and its probed candidates, which varies with agent co-location but is likewise independent of collective size.

\paragraph{Load Balancing and Scheduling.}

We run 200 randomly selected tasks at a concurrency of five, yielding a 1:1 ratio of tasks to workers per task to simulate a loaded collective. We configure \panda with a probe ratio $d = 2$ and a FIFO scheduling policy. To make the comparison fair with respect to concurrency, each baseline is fronted with a server that accepts and distributes concurrent requests across its underlying architecture. We provide results in Table~\ref{tab:hotpotqa}.

\begin{table}[h]
\centering
\caption{Performance on HotPotQA benchmark without failures.}
\label{tab:hotpotqa}

\setlength{\tabcolsep}{5pt}
\renewcommand{\arraystretch}{1.25}
\small
\begin{tabular}{@{}lcccccc@{}}
\toprule
& \multicolumn{3}{c}{Complete Knowledge} & \multicolumn{3}{c}{Partial Knowledge} \\
\cmidrule(lr){2-4} \cmidrule(lr){5-7}
System & EM (\%) & F1 (\%) & Total RT (s) & EM (\%) & F1 (\%) & Total RT (s) \\
\midrule
Magentic-One  & 42.5 & 48.52 & 851.1 & 25.5 & 29.99 & 1181.7 \\
AgentNet  & 67.5 & 77.33 & 840.4 & 43.5 & 51.37 & 1588.4 \\
Internet of Agents & \textbf{72.5} & \textbf{81.26} & 2182.1 & \textbf{62.0} & \textbf{68.07} & 3949.1 \\
\midrule
\pandastar & 66.5 & 74.11 & \underline{464.5} & 45.5 & 54.60 & \underline{594.9} \\
\pandachain & 64.5 & 75.54 & \textbf{462.5} & 47.0 & 53.18 & \textbf{452.9} \\
\pandamesh & \underline{70.0} & \underline{79.39} & 714.0 & \underline{56.0} & \underline{65.90} & 1038.4 \\
\bottomrule
\end{tabular}
\end{table}

\panda significantly speeds up performance with respect to our baselines. \pandastar and \pandachain are roughly $2\times$ faster than Magentic-One and AgentNet improving EM and F1 over Magentic-One and comparable with AgentNet. IoA records higher accuracy than \pandastar and \pandachain but incurs $5$-$8 \times$ runtime. \pandamesh is faster than Magentic-One and AgentNet and achieves better accuracy metrics especially in the partial knowledge case. \pandamesh essentially matches IoA in terms of EM and F1 but with a $3$-$4\times$ faster runtime. Overall we demonstrate that \panda can handle a high throughput of tasks while maintaining state-of-the-art EM and F1. 

\subsection{Task Completion with Failures}
\label{sec:eval_fail}

\paragraph{Infrastructure Failures.} 
For each task, we ensure that at least one agent crashes during execution. Since the number of steps a task will take is not known, we set a high per-step crash probability of $p=0.5$, with a backstop guaranteeing at least one crash per task. We present the results in Table~\ref{tab:crash_all}. 

\begin{table}[h]
\centering
\caption{Performance on HotPotQA under infrastructure failures.}
\label{tab:crash_all}
\setlength{\tabcolsep}{3.5pt}
\renewcommand{\arraystretch}{1.25}
\small
\resizebox{\textwidth}{!}{%
\begin{tabular}{@{}lcccccccc@{}}
\toprule
& \multicolumn{4}{c}{Complete Knowledge} & \multicolumn{4}{c}{Partial Knowledge} \\
\cmidrule(lr){2-5} \cmidrule(lr){6-9}
System & Completion (\%) & F1 (\%) & Comp. RT (s) & Failed RT (s) 
       & Completion (\%) & F1 (\%) & Comp. RT (s) & Failed RT (s)  \\
\midrule
Magentic-One & 0.00 & 0.00 & -- & \underline{8.14} & 0.00 & 0.00 & -- & \textbf{8.92}  \\
AgentNet & 0.00 & 0.00 & -- & \textbf{5.08} & 0.00 & 0.00 & -- & \underline{15.91}  \\
Internet of Agents & 0.00 & 0.00 & -- & 100.63 & 0.00 & 0.00 & -- & 122.63  \\
\midrule
\pandastar  & \textbf{100.0} & \textbf{72.33} & \underline{28.04} & -- & \textbf{100.0} & \textbf{68.10} & 39.42 & --  \\
\pandachain & \textbf{100.0} & 68.23 & \textbf{27.29} & -- & \textbf{100.0} & \underline{60.43} & \underline{32.23} & --  \\
\pandamesh  & \textbf{100.0} & \underline{71.80} & 61.57 & -- & \textbf{100.0} & 57.13 & \textbf{31.87} & --  \\
\bottomrule
\end{tabular}}
\end{table}

None of the baselines complete any task under agent crashes. In contrast, all three \panda topologies sustain high accuracy and reasonable per-task runtimes. We note that Magentic-One and AgentNet fail quickly, whereas IoA has no mechanism to detect a crash until a subtask times out, inflating its failure latency. We define a failed task as a task that did not complete.

\paragraph{Orchestration Failures.} 
For each task, one agent is faulty at the start of execution, always returning an unrelated search result. In AgentNet, IoA, and \panda, healthy replacements exist in the registry and can be discovered; Magentic-One has no mechanism to swap a faulty agent and is stuck with the misconfigured one. A task is \emph{correct} if F1 $> 0$, and \emph{incorrect} if F1 $= 0$. We present the results in Table~\ref{tab:potato}.

\begin{table}[h]
\centering
\caption{Performance on HotPotQA under orchestration failures.}
\label{tab:potato}

\setlength{\tabcolsep}{3.5pt}
\renewcommand{\arraystretch}{1.25}
\small
\resizebox{\textwidth}{!}{%
\begin{tabular}{@{}lcccccccc@{}}
\toprule
& \multicolumn{4}{c}{Complete Knowledge} & \multicolumn{4}{c}{Partial Knowledge} \\
\cmidrule(lr){2-5} \cmidrule(lr){6-9}
System & EM (\%) & F1 (\%) & Correct RT (s) & Incorrect RT (s)
       & EM (\%) & F1 (\%) & Correct RT (s) & Incorrect RT (s) \\
\midrule
Magentic-One       & 45.00 & 52.31 & 28.24 & 59.58 & 6.50 & 6.83 & 18.01 & 40.07 \\
AgentNet           & 1.00 & 1.56 & 47.90 & 24.57 & 22.50 & 27.51 & 72.90 & 50.59 \\
Internet of Agents & 24.50 & 27.34 & 131.40 & 192.78 & 11.00 & 12.51 & 57.87 & 278.40 \\
\midrule
\pandastar  & \underline{64.00} & \underline{71.19} & \textbf{13.93} & \underline{23.69} & 49.50 & 59.05 & \textbf{15.91} & 29.42 \\
\pandachain & 58.50 & 69.33 & 19.69 & \textbf{23.66} & \underline{53.50} & \underline{62.89} & 22.73 & \underline{25.32} \\
\pandamesh  & \textbf{69.50} & \textbf{78.86} & \underline{15.73} & 26.59 & \textbf{54.00} & \textbf{66.14} & \underline{22.70} & \textbf{24.51} \\
\bottomrule
\end{tabular}}
\end{table}

Under faults, \panda outperforms the baselines across all metrics. Its dedicated swap protocol replaces the faulty agent with a healthy one, allowing the task to complete correctly and efficiently. Complete knowledge provides redundancy that improves EM and F1 for Magentic-One and IoA; however, for AgentNet, the added confidence in capability causes the faulty agent to return incorrect answers with higher probability.

\subsection{PANDA Governance}
\label{sec:eval_gov}

We study how trust propagates in \panda. Our experiment proceeds in two stages. In the \emph{join} stage (60\,s), 1000 agents enter the collective, each collecting $t=2$ certificates from existing agents that verify and vouch for its identity. Joins are uniformly distributed across the stage, and certifiers are drawn uniformly at random from the agents active in the collective. In the \emph{churn} stage (30\,s), agents join and leave the collective, each at a rate of one per second following a Poisson process.

We evaluate the \emph{self} and \emph{genesis} anchors at depths $n =3, 5, 7, \infty$ and measure \emph{trusted recall}: the fraction of alive agents that each agent trusts. We visualize the underlying certificate graph, in which each node is an agent and each edge $(A_i, A_j)$ indicates that $A_i$ and $A_j$ have mutually signed each other's certificates. Results are shown in Figure~\ref{fig:trust_all}; (see Appendix~\ref{app:more_gov_experiments} for additional results). All configurations achieve trusted precision, the fraction of trusted peers that are actually alive, $>99\%$.

\begin{figure}[h]
    \centering
    \begin{subfigure}[t]{0.355\linewidth}
        \centering
        \includegraphics[width=\linewidth]{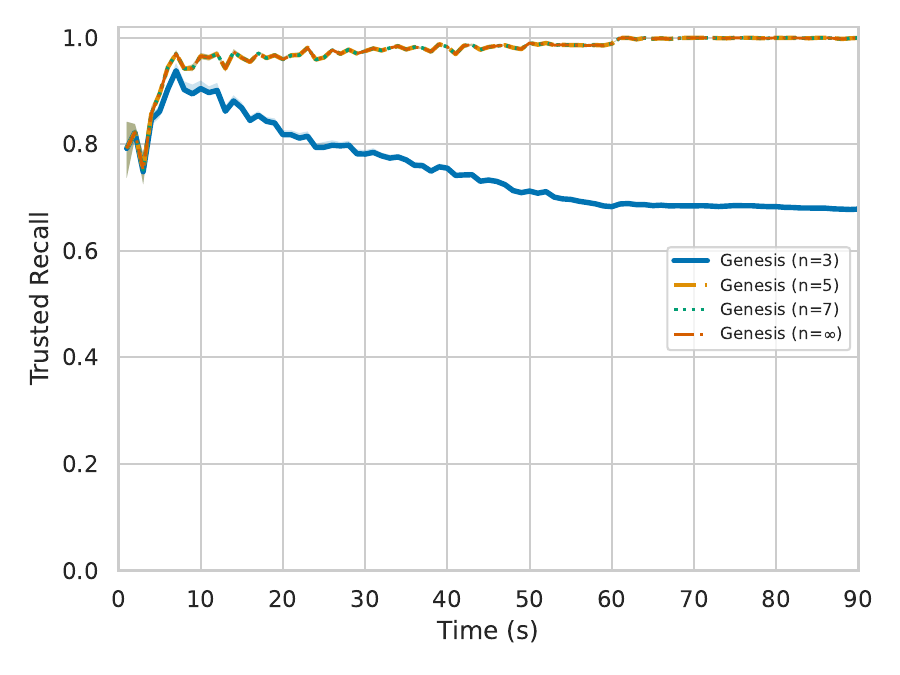}
        \caption{Genesis Anchor ($t=2$)}
        \label{fig:root_web_recall}
    \end{subfigure}
    \hfill
    \begin{subfigure}[t]{0.355\linewidth}
        \centering
        \includegraphics[width=\linewidth]{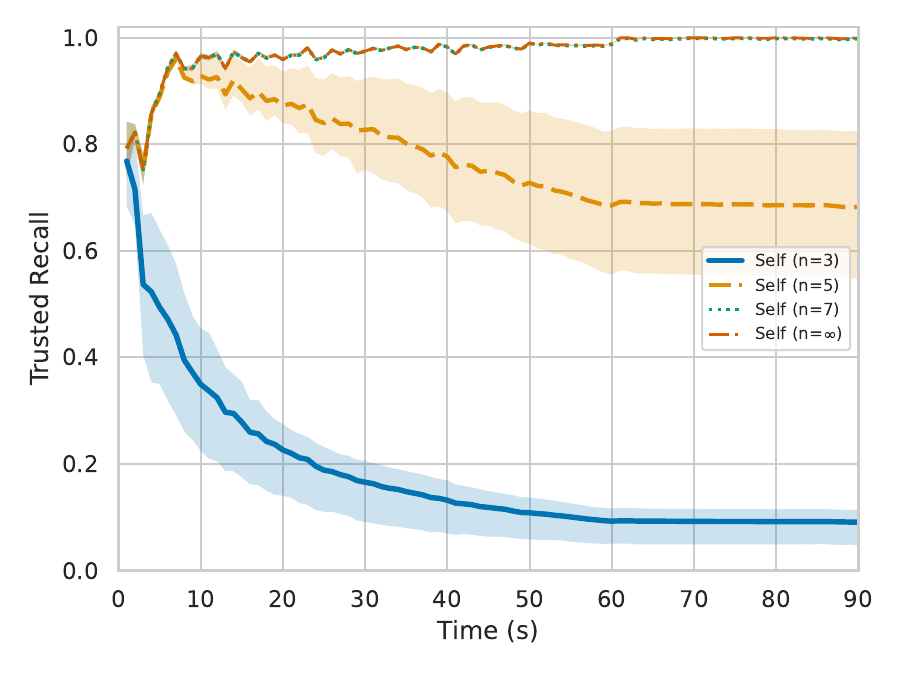}
        \caption{Self Anchor ($t=2$)}
        \label{fig:self_web_recall}
    \end{subfigure}
    \hfill
    \begin{subfigure}[t]{0.265\linewidth}
        \centering
        \includegraphics[width=\linewidth]{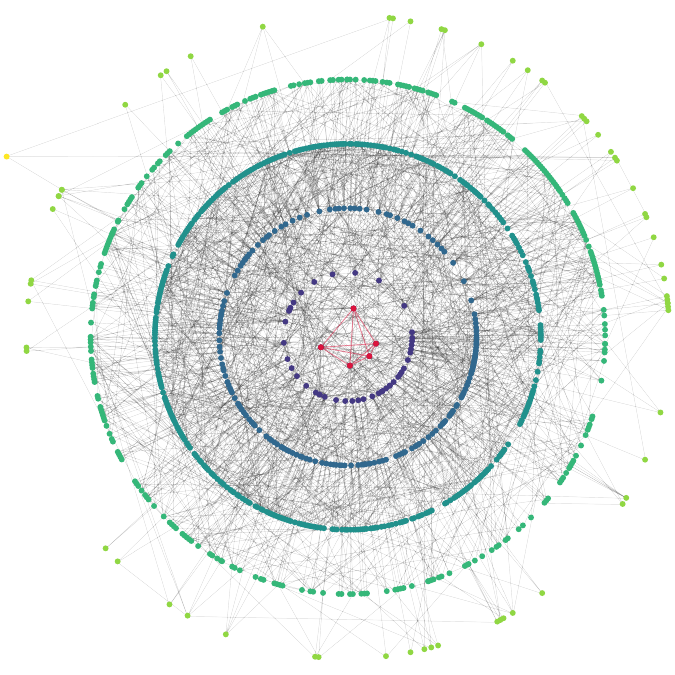}
        \caption{Certificate Graph ($t=2$)}
        \label{fig:web_graph}
    \end{subfigure}

    \caption{Average trusted recall in a collective where each agent requires two certificates to join with five genesis agents. Shaded regions show the IQR of trusted recall across agents.}
    \label{fig:trust_all}
\end{figure}

Under the genesis anchor, coverage is strong: even at $n=3$, trusted recall reaches nearly $70\%$. Genesis anchors have essentially no IQR, since all genesis nodes are mutually connected. Self anchoring is weaker at lower $n$ but still recovers a trusted set of nearly 100 agents at $n=3$. Overall, trusted set size remains stable under churn, and even restrictive trust schemes build substantial trusted sets.

\section{Conclusion}

In this work, we presented \panda, a decentralized architecture for large-scale LLM-based multi-agent systems, addressing four requirements that existing architectures satisfy either partially or not at all: scalability, fault tolerance, flexible orchestration, and decentralized governance. \panda distributes agents and tasks across the infrastructure, preserves task execution despite orchestration and infrastructure failures, and supports multiple planning and execution models. Its web-of-trust governance model enables agents from different administrative domains to establish trust without relying on a single central authority. Our evaluation demonstrates that \panda scales efficiency while maintaining task utility, recovering from failures, and supporting diverse orchestration patterns, providing a foundation for building open, resilient, and governable multi-agent ecosystems.
% \newpage
\subsubsection*{Acknowledgments}
This material is based upon work supported by the National Science Foundation Graduate Research Fellowship Program under Grant No. DGE-2439018. Any opinions, findings, and conclusions or recommendations expressed in this material are those of the author(s) and do not necessarily reflect the views of the National Science Foundation. This research was also supported by funding from Google.

\bibliography{aa-references}
\bibliographystyle{arxiv}

\appendix
\section{Team Assembly and Repair}

Team formation and maintenance are fundamental components of \panda. In this appendix we formalize initial team assembly and team repair via the \texttt{swap} protocol.

\subsection{Assembly Protocol}

\label{app:asm_protocol}
Determining whether an agent can fulfill a capability relies on a matching primitive $\texttt{match}(c, a) \in \{0, 1\}$, which can be realized in various ways, including direct string comparison or thresholded semantic scoring. Given a set of capabilities $\mathcal{C}$, a redundancy $r(c) \in \mathbb{Z}^+$ and a trusted set of agents $\mathcal{A}$, we construct a matching $\phi : \mathcal{C} \to 2^{\mathcal{A}}$ of agents to capabilities, assigning each capability $c$ a set $\phi(c) \subseteq \mathcal{A}$ where $|\phi(c)| = r(c)$ agents. We require $\phi$ to be \emph{valid}: every assigned agent must possess the corresponding capability, i.e., $\forall a \in \phi(c): \texttt{match}(c, a) = 1$. Among valid matchings, we consider two strategies. A \emph{dedicated} matching assigns each capability its own distinct pool, with no agent shared across capabilities:
\begin{equation}
    \phi(c_i) \cap \phi(c_j) = \emptyset \quad \forall\, c_i \neq c_j \in \mathcal{C}.
\end{equation}
This maximizes parallelism at the cost of recruiting $\sum_{c \in \mathcal{C}} r(c)$ agents. A \emph{consolidated} matching instead minimizes the total number of distinct agents recruited:
\begin{equation}
    \min_{\phi} \;\Big| \bigcup_{c \in \mathcal{C}} \phi(c) \Big|.
\end{equation}
Computing a minimal consolidated matching is an instance of set cover, a known NP-hard problem \citep{nphard}; we therefore opt for an approximate greedy solution instead.

The assembly procedure proceeds as follows. The entrypoint, the agent to which the request was sent, first consults its local registry. If it can assemble a valid matching $\phi$ from registered agents, team assembly completes locally. Otherwise, the entrypoint broadcasts a request for the unmatched capabilities, and any trusted agent able to fulfill one may respond and join the team.

\subsection{Swap Protocol}
\label{app:swap_proto}
When an agent $a_i$ that serves some capability $c$ crashes or is declared failed, we must remove it from the team and replace its capabilities. We refer to this as the \texttt{swap} protocol, denoted $\texttt{swap}(a_i, c)$. If, for some capability $c$, sufficient redundancy $r(c)$ remains such that $|\phi(c) \setminus \{a_i\}| > 0$, we can rely on the other agents $a_j \in \phi(c) \setminus \{a_i\}$ covering that capability: any subtasks previously assigned to $a_i$ are reassigned to some $a_j$, and $a_i$ is evicted from the team: $\phi(c) \gets \phi(c) \setminus \{a_i\}$ so no further subtasks are assigned to it.

If $|\phi(c)|$ ever reaches zero, a replacement must be found. Similar to team assembly, the agent that detected the fault first searches its registry; if successful, it adds the found agent $a_k$ to the team updating $\phi(c) \gets \{a_k\}$. Otherwise, it broadcasts the request and awaits a response, updating $\phi(c)$ when one arrives.

\section{Fault-Tolerant Orchestration Details} 
\label{app:topos}

\panda's architecture naturally accommodates a variety of execution patterns. In this appendix, we explore in detail three general execution topologies into which \panda teams can be shaped, and under which the majority of state-of-the-art multi-agent systems fall. For each topology, we design a planning strategy that produces a structured plan, decomposing a task into a graph of subtasks, each corresponding to a single capability. We then extend each design to be fault-tolerant, where the step-based plan facilitates efficient recovery. 

\begin{table}
\centering
\caption{Summary of notation.}
\label{tab:notation}
\begin{tabular}{@{}r@{\quad}l@{}}
\toprule
\multicolumn{2}{@{}l}{\textit{Task \& Team}} \\
\midrule
$\mathcal{T}$ & User task \\
$\mathcal{C}$ & Required capabilities \\
$\mathcal{A'}$ & Team of agents \\
$r(c)$ & Redundancy: agents recruited per capability \\
$\phi(c)$ & Pool of $r(c)$ agents matched to capability $c$ \\
\addlinespace
\multicolumn{2}{@{}l}{\textit{Plan}} \\
\midrule
$\mathcal{P}$ & Plan (DAG of subtasks) \\
$\mathcal{V}, \mathcal{E}$ & Subtasks and dependencies \\
$\mathrm{cap}(v)$ & Capability required by subtask $v$ \\
$\delta(v)$ & Agent assigned to subtask $v$ \\
\addlinespace
\multicolumn{2}{@{}l}{\textit{State \& Recovery}} \\
\midrule
$\mathcal{S}$ & Global task state $(\mathcal{P}, \mathcal{D}, \mathcal{O}, \mathcal{F})$ \\
$\mathcal{D}$ & Completed subtasks \\
$\mathcal{O}$ & Outputs of completed subtasks \\
$\mathcal{F}$ & In-flight subtasks \\
\bottomrule
\end{tabular}
\end{table}

\subsection{Problem Formulation}
In each topology, the goal is to solve a user task $\mathcal{T}$ given the tuple $(\mathcal{T}, \mathcal{C}, r, \mathcal{A'}, \phi)$, where $\mathcal{C}$ is the set of required capabilities, $r : \mathcal{C} \to \mathbb{Z}^{+}$ is the redundancy mapping, $\mathcal{A'}$ is the team of agents, and $\phi : \mathcal{C} \to 2^{\mathcal{A'}}$ is the matching assigning each capability $c$ to a pool of $|\phi(c)| = r(c)$ capable agents (Appendix~\ref{app:asm_protocol}).
To solve $\mathcal{T}$, we first generate a plan $\mathcal{P}$, represented as a directed acyclic graph (DAG) of subtasks, adapting graph-structured reasoning to a multi-agent setting \citep{GoT}. The vertices $\mathcal{V}$ represent subtasks and the edges $\mathcal{E}$ represent dependencies between them. Each vertex $v_i \in \mathcal{V}$ is assigned a capability $\mathrm{cap}(v_i) \in \mathcal{C}$ and a capable agent $\delta(v_i) \in \phi(\mathrm{cap}(v_i))$:
\begin{equation}
    \mathrm{cap} : \mathcal{V} \to \mathcal{C}, \qquad \delta : \mathcal{V} \to \mathcal{A'}, \qquad \delta(v_i) \in \phi(\mathrm{cap}(v_i)).
\end{equation}
Once the plan is generated, execution proceeds according to the specified topology. Across all topologies, if an agent fails, we invoke the $\texttt{swap}(a, c)$, where a crashed agent $a$ is replaced by another agent $a'$ that fulfills the same capability $c$. If redundancy permits ($r(c) > 1$), a replacement is available immediately; otherwise, a new agent is added dynamically by searching for $c$ via the \texttt{swap} protocol (Appendix~\ref{app:swap_proto}).

\subsection{Star Topology}

\paragraph{Design.} The star topology is common in multi-agent systems, where all planning is handled by an orchestrator at the center of the star \citep{magentic-one, owl}. In \pandastar, the orchestrator receives the task, $\mathcal{T}$, and generates the plan, $\mathcal{P}$. It begins by executing all dependency-free subtasks in parallel. As dependencies are resolved, the orchestrator optionally synthesizes intermediate results and dispatches each newly enabled subtask to its assigned agent, along with the relevant context from completed dependencies. The orchestrator maintains the global task state $\mathcal{S} = (\mathcal{P}, \mathcal{D}, \mathcal{O}, \mathcal{F})$, where $\mathcal{P}$ is the plan, $\mathcal{D} \subseteq \mathcal{V}$ is the set of completed subtasks, $\mathcal{O}$ are the saved outputs, and $\mathcal{F} \subseteq \mathcal{V}$ are the in-flight subtasks. Once all subtasks are fulfilled, the result is returned to the user.

\paragraph{Replanning.}
In order to remain tolerant to failures, the orchestrator can replan. After each subtask completes, the orchestrator considers the new output together with the current state $\mathcal{S}$ and decides whether to replan. If so, it generates a revised plan $\mathcal{P}'$ while holding the already-completed subtasks $\mathcal{D}$ fixed. It then reconciles the in-flight subtasks $\mathcal{F}$ -- any no longer required under $\mathcal{P}'$ are canceled, while the rest continue running.

\paragraph{Fault Tolerance.} There are two components of the star topology that can fail, the orchestrator or the workers. Worker infrastructure failures are straightforward: the orchestrator uses \panda's failure detector to detect a crashed agent $a \in \phi(c)$ and swap it with another agent $a'$ using the \texttt{swap} protocol. If the crashed agent was actively executing a subtask $v_i$ ($\delta(v_i) = a$), the orchestrator reissues it, updating $\delta(v_i) = a'$; otherwise, execution proceeds as normal with a fully restored team. Orchestration failures, such as repeated bad responses, are reasoned about after each subtask completes. An orchestrator can elect to invoke the swap protocol for an agent that repeatedly fails to give a reasonable answer. 

We next consider failure of the orchestrator itself. The simplest response is to restart the task from scratch, however, this discards all completed work. Instead, the orchestrator, $a_{\text{orch}}$, can checkpoint its execution state $\mathcal{S}$ to one or more peers as it progresses, so that upon failure a replacement orchestrator can resume from the latest checkpoint rather than from the beginning. The entrypoint or a neighboring orchestrator are natural candidates for holding these checkpoints. Whichever peer holds the checkpoint detects the orchestrator's failure and restarts the task from the checkpointed state at a new orchestrator, $o' = \texttt{swap}(a_{\text{orch}}, \text{``orchestration''})$.

\paragraph{Redundant Execution.}

For even stronger fault tolerance, at the cost of additional computation, a subtask $v$ can be executed redundantly whenever its capability pool contains spare agents ($r(\mathrm{cap}(v)) > 1$). Rather than assigning $v$ to a single agent, we assign it to $n \leq r(\mathrm{cap}(v))$ agents drawn from $\phi(\mathrm{cap}(v))$, accept the first response, and cancel the remaining executions. Because the subtask completes as soon as any one agent succeeds, it tolerates up to $n - 1$ failures without triggering a swap or incurring recovery latency.

\subsection{Chain Topology}

\paragraph{Design.} The chain topology consists of agents that communicate without a centralized point of control. Agents are arranged in a dynamic chain in which each agent communicates only with its immediate predecessor and successor. \panda applies dynamic, decentralized planning to this topology. A randomly chosen agent $a_0$ first generates a preliminary plan $\mathcal{P}$. Without a central point of control, coordinating concurrent subtasks quickly becomes intractable, so $\mathcal{P}$ is restricted to a sequential plan $v_1, \dots, v_n$ -- which our DAG formulation handles gracefully, as a sequence is trivially a DAG. $a_0$ then forwards $\mathcal{P}$, along with the task, to $\delta(v_1)$, the agent assigned to its first subtask. This repeats down the chain: each agent $a = \delta(v_i)$ fulfills its subtask $v_i$ and forwards the plan to $\delta(v_{i+1})$. Once all subtasks are complete, the final agent returns the result.

\paragraph{Replanning.}
Upon receiving a subtask, an agent $a = \delta(v_i)$ for the pending subtask $v_i$ evaluates it. %as shown in Appendix \ref{app:chain_replan}. 
If unsatisfactory, $a$ may trigger a \emph{replan}, producing a revision $\mathcal{P}' = v'_i, \dots, v'_m$ in place of the remaining plan. If the new pending subtask is its own ($\delta(v'_i) = a$), $a$ executes $v'_i$ and forwards the plan to $\delta(v'_{i+1})$; otherwise, $a$ forwards the revised plan to $\delta(v'_i)$ to fulfill. The number of replans is capped to guarantee termination. This decentralized replanning lets agents progressively rewrite the plan to better suit the team's capabilities. 

\paragraph{Fault Tolerance.}
Consider three consecutive agents in the chain, $a_{i-1}, a_i, a_{i+1}$, assigned to subtasks $v_{i-1}, v_i, v_{i+1}$ respectively. When $a_{i-1}$ completes $v_{i-1}$, it forwards the plan and its output to $a_i$ but retains a copy of what it sent. Orchestration failures, such as misconfigured agents can be detected and swapped during the replanning step. The challenge is detecting and recovering from infrastructure failures of $a_i$ without global coordination.

We introduce a \texttt{DONE} acknowledgment for this purpose. Once $a_i$ has fulfilled $v_i$ and forwarded its result to $a_{i+1}$, it sends \texttt{DONE} back to $a_{i-1}$ in the background, signaling that responsibility has passed downstream. If $a_{i-1}$ declares $a_i$ failed (via \panda's failure detector) before receiving this \texttt{DONE}, it invokes $\texttt{swap}(a_i, \mathrm{cap}(v_i))$ to obtain a replacement $a'_i$ and reissues $v_i$ -- using its retained copy of the handoff state -- to $a'_i$. Because each agent retains its handoff until acknowledged, recovery requires only local state at the immediate predecessor. 

A duplicate chain can arise in an edge case: $a_i$ forwards to $a_{i+1}$ but fails before sending \texttt{DONE} to $a_{i-1}$, causing $a_{i-1}$ to reissue $v_i$ even though the work has already progressed downstream. To prevent this, each reissue carries a unique, monotonically increasing \emph{generation} number. Every downstream agent records the highest generation it has seen and terminates any message belonging to a lower generation, so the stale chain is killed while the reissued one proceeds.

\paragraph{Redundant Execution.}

As with the star, we can trade computational cost for increased fault tolerance: rather than dispatching $v_i$ to a single successor, $a_{i-1}$ assigns it to $n \leq r(\mathrm{cap}(v_i))$ agents from $\phi(\mathrm{cap}(v_i))$ and, upon receiving a \texttt{DONE} from any one, cancels the rest. The subtask then tolerates up to $n - 1$ failures without invoking the \texttt{swap} protocol.

\subsection{Mesh Topology} 

\paragraph{Design.}

The mesh topology consists of a team of agents arranged in a fully connected mesh, each able to communicate with all others. The team's first step is to agree on a plan $\mathcal{P}$, represented as a DAG. To reach agreement efficiently, we first select a subset of collaborators $\mathcal{K} \subseteq \mathcal{A'}$ that covers every required capability:
\begin{equation}
    \mathcal{K} = \{\, a_c : a_c \in \phi(c) \,\}_{c \in \mathcal{C}}
\end{equation}
Restricting planning to $\mathcal{K}$ rather than the full team keeps the process efficient while ensuring all capabilities are represented. Each collaborator $a_i \in \mathcal{K}$ generates a candidate plan $\mathcal{P}_i$. Every plan is broadcast to the other collaborators, who rank them via ranked-choice voting. The resulting scores are broadcast so that every collaborator holds the same set of scores and thus a consistent view of the rankings. Each agent then independently eliminates the bottom $j$ proposals -- arriving at the same outcome from the shared scores -- and feedback is returned to the authors of the surviving plans, who revise them. This propose--rank--revise cycle repeats until a single consensus plan $\mathcal{P}$ remains. % See %Appendix~\ref{app:mesh_plan} for more details.

Once a plan is agreed upon, the team executes it. The plan is first broadcast to the entire team, including redundant agents, and every subtask without dependencies is assigned. Fulfillment of a subtask $v$ proceeds as follows. If multiple agents are assigned to its capability ($|\phi(\mathrm{cap}(v))| > 1$), they jointly produce a result through multi-agent debate \citep{multi_agent_debate}: each agent proposes an answer and critiques the others over successive rounds until they converge on an agreed output $o_v$. Otherwise, a single agent produces $o_v$ directly. When $v$ completes, $o_v$ is broadcast to the team, and any subtask whose dependencies are now satisfied may begin. Each agent maintains the global task state $\mathcal{S} = (\mathcal{P}, \mathcal{D}, \mathcal{O}, \mathcal{F})$. Once all subtasks are fulfilled, the result is returned to the user.

\paragraph{Replanning.}
After each subtask output is fulfilled, the collaborators $\mathcal{K}$ vote on whether to replan. If a supermajority (more than $2/3$) votes in favor, each agent that voted yes proposes a revised plan $\mathcal{P}'_i$, holding the already-completed subtasks $\mathcal{D}$ fixed. The same propose--rank--revise cycle then runs over these proposals to produce the new plan $\mathcal{P}'$. Once $\mathcal{P}'$ is broadcast to the team, each agent independently cancels any in-flight subtask $v_f \in \mathcal{F}$ no longer required under $\mathcal{P}'$, while the rest continue running.

\paragraph{Fault Tolerance.}

Nodes can fail at two points: planning or execution. If a collaborator $a_f \in \mathcal{K}$ responsible for capability $c_f$ fails during planning, we perform $\texttt{swap}(a_f, c_f)$; if $a_f$'s proposal is still in contention, it is removed from the running. This can never eliminate the last remaining plan, since execution begins as soon as a single plan remains. If instead the failed node was a redundant agent (not in $\mathcal{K}$), the failure can be ignored, as it does not affect the team's ability to plan.

If an agent fails during execution, there are two cases. If the failed agent belongs to a redundant pool ($|\phi(c)| > 1$) and is actively participating in the debate for a subtask $v$ with $\mathrm{cap}(v) = c$, the remaining agents in $\phi(c)$ simply drop it from the debate and continue to produce $o_v$; if that pool is idle at the time of failure, the failure can be ignored. If instead the capability has no redundancy ($|\phi(c)| = 1$), we recover via $\texttt{swap}(a, c)$.

\paragraph{Redundant Execution.}

As with the previous two topologies, the mesh topology can execute redundantly. However, because the mesh already uses all available redundancy for multi-agent debate, redundant execution is offered as an alternative rather than an additive extension. Multi-agent debate optimizes for accuracy at the cost of additional rounds of communication; redundant execution instead has all agents in $\phi(\mathrm{cap}(v))$ execute the subtask $v$ in parallel and accepts the first response, cancelling the rest. This tolerates up to $|\phi(\mathrm{cap}(v))| - 1$ failures without triggering a swap or incurring recovery latency, trading debate's accuracy gains for lower latency.
\section{Formal Verification}

We formally specify \panda's core protocols in TLA\textsuperscript{+}, modeling peer gossip, infrastructure failure detection, failure recovery, self-refutation, team assembly, and the swap protocol. Load-aware candidate selection is abstracted as nondeterministic choice, so safety holds under any selection policy. Using TLC, we verify three properties: every completed task has all required capabilities assigned (safety), no agent believes it belongs to a team it was not assigned to (safety), and every task eventually completes or fails under standard fairness assumptions (liveness).
\section{Additional Governance Experiments}
\label{app:more_gov_experiments}

In this appendix we expand our study of how trust propagates in \panda. As in Section~\ref{sec:eval_gov}, each experiment proceeds in two stages. In the \emph{join} stage, agents enter the collective, each exchanging $t$ mutual certificates with existing agents that verify its identity. In the \emph{churn} stage, agents join and leave the collective, each at a rate of one per second following a Poisson process.

For each policy, we evaluate the \emph{self} and \emph{genesis} anchors at depths $n \in \{3, 5, 7, \infty\}$, measure \emph{trusted recall}, and visualize the underlying certificate graph (as defined in Section~\ref{sec:eval_gov}). All configurations achieve trust precision $> 99\%$.

\subsection{Varying the Number of Certifiers}

\begin{figure}[h]
    \centering
    \begin{subfigure}[t]{0.355\linewidth}
        \centering
        \includegraphics[width=\linewidth]{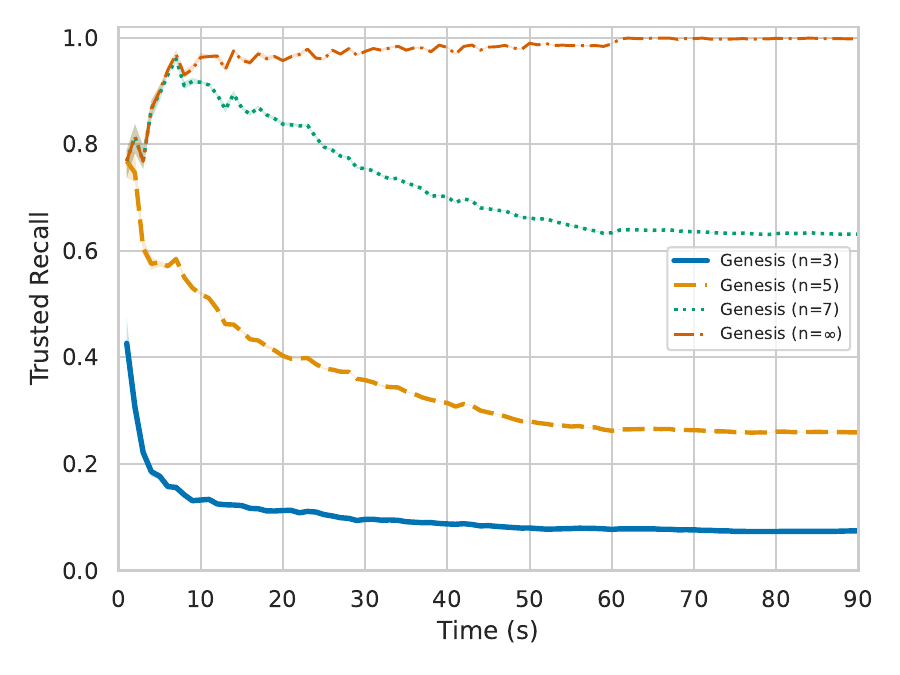}
        \caption{Genesis Anchor ($t=1$)}
        \label{fig:root_tree_recall}
    \end{subfigure}
    \hfill
    \begin{subfigure}[t]{0.355\linewidth}
        \centering
        \includegraphics[width=\linewidth]{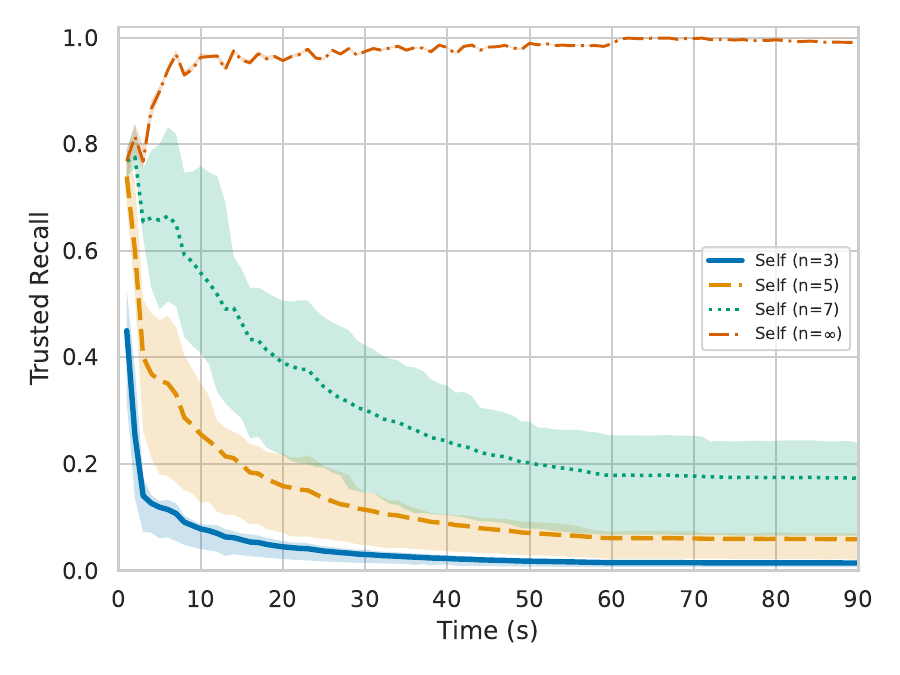}
        \caption{Self Anchor ($t=1$)}
        \label{fig:self_tree_recall}
    \end{subfigure}
    \hfill
    \begin{subfigure}[t]{0.265\linewidth}
        \centering
        \includegraphics[width=\linewidth]{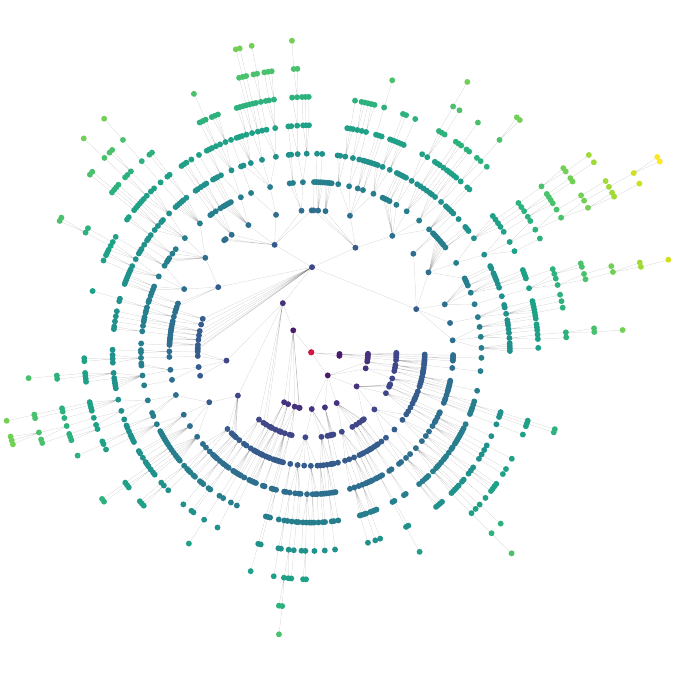}
        \caption{Certificate Graph ($t=1$)}
        \label{fig:tree_graph}
    \end{subfigure}

    \caption{Average trusted recall in a 1000 agent collective where each agent requires one certificate to join with one genesis agent. Shaded regions show the IQR of trusted recall across agents.}
    \label{fig:tree_all}
\end{figure}

\begin{figure}[h]
    \centering
    \begin{subfigure}[t]{0.355\linewidth}
        \centering
        \includegraphics[width=\linewidth]{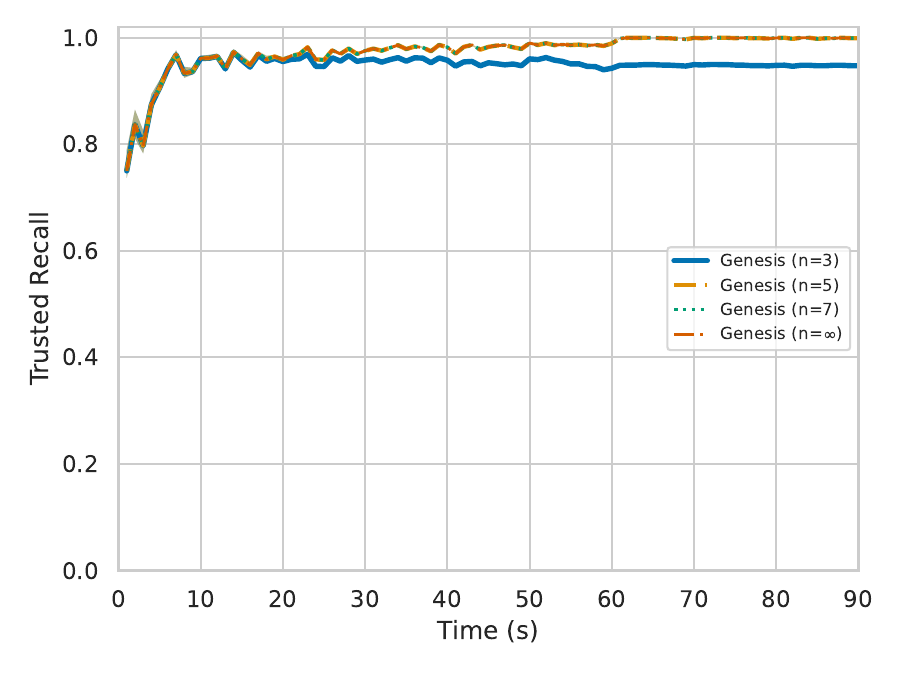}
        \caption{Genesis Anchor ($t=3$)}
        \label{fig:root_web3_recall}
    \end{subfigure}
    \hfill
    \begin{subfigure}[t]{0.355\linewidth}
        \centering
        \includegraphics[width=\linewidth]{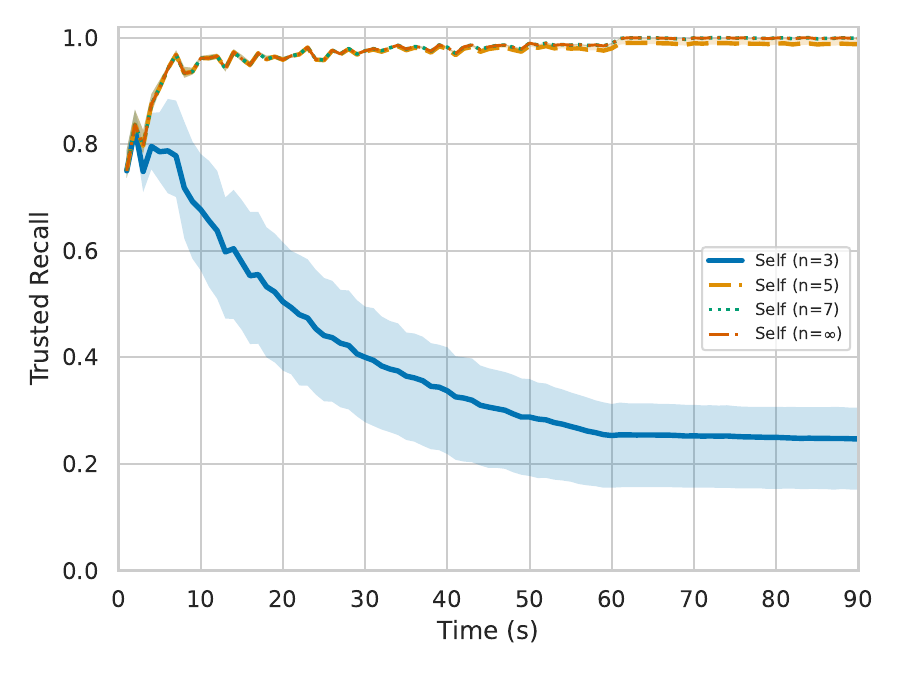}
        \caption{Self Anchor ($t=3$)}
        \label{fig:self_web3_recall}
    \end{subfigure}
    \hfill
    \begin{subfigure}[t]{0.265\linewidth}
        \centering
        \includegraphics[width=\linewidth]{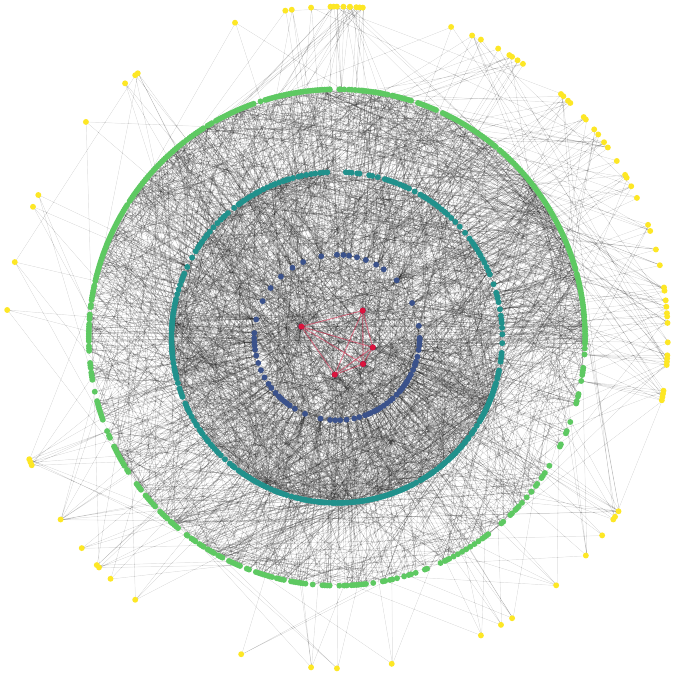}
        \caption{Certificate Graph ($t=3$)}
        \label{fig:web3_graph}
    \end{subfigure}

    \caption{Average trusted recall in a 1000 agent collective where each agent requires three certificates to join with five genesis agents. Shaded regions show the IQR of trusted recall across agents.}
    \label{fig:web3_all}
\end{figure}

We present results for $t=1$ and $t=3$ on a collective size of $1000$ in Figures~\ref{fig:tree_all} and~\ref{fig:web3_all}, respectively. At $t=1$, the reduced certification requirement yields a significantly less connected graph, but coverage remains strong under the genesis anchor: even at $n=3$, each agent trusts nearly 100 peers. Self anchoring is weaker but scales with depth, achieving near-perfect recall at $n=\infty$ because the trust graph is connected. While fewer connections reduce the trusted set overall, substantial trusted sets can still be accrued when the collective is sufficiently large. At $t=3$, trusted recall improves relative to $t=2$ (Section~\ref{sec:eval_gov}), though the gain is smaller than the jump from $t=1$ to $t=2$, suggesting diminishing returns. Together these results suggest $t=2$ as a reasonable default -- also the choice made by the Debian developers group \citep{debian}. In both cases, trusted set size remains stable under churn.

\subsection{Larger Collectives}

\begin{figure}[h]
    \centering
    \begin{subfigure}[t]{0.355\linewidth}
        \centering
        \includegraphics[width=\linewidth]{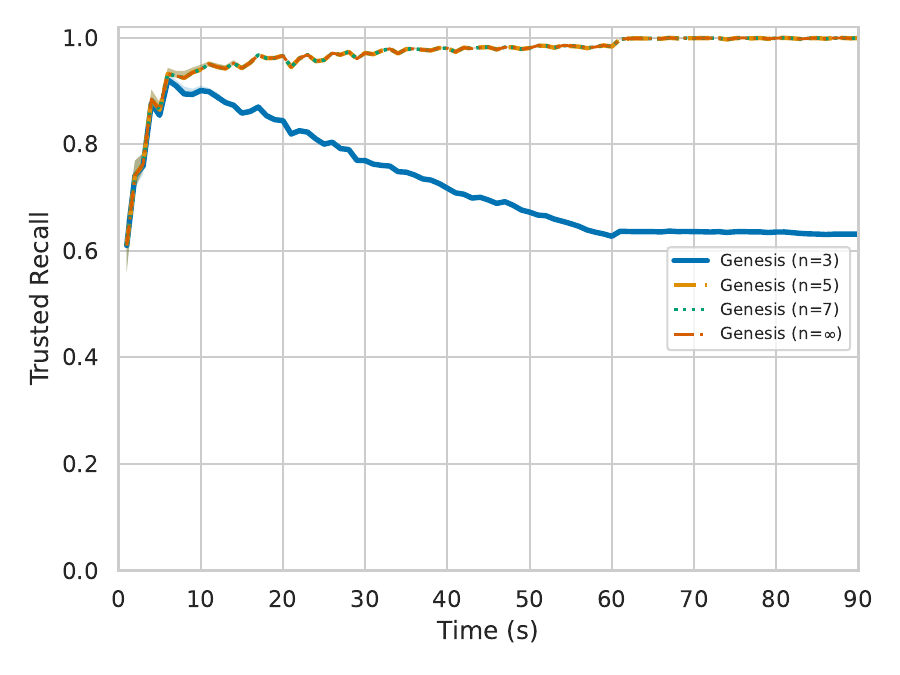}
        \caption{Genesis Anchor ($t=2$)}
        \label{fig:root_web2k_recall}
    \end{subfigure}
    \hfill
    \begin{subfigure}[t]{0.355\linewidth}
        \centering
        \includegraphics[width=\linewidth]{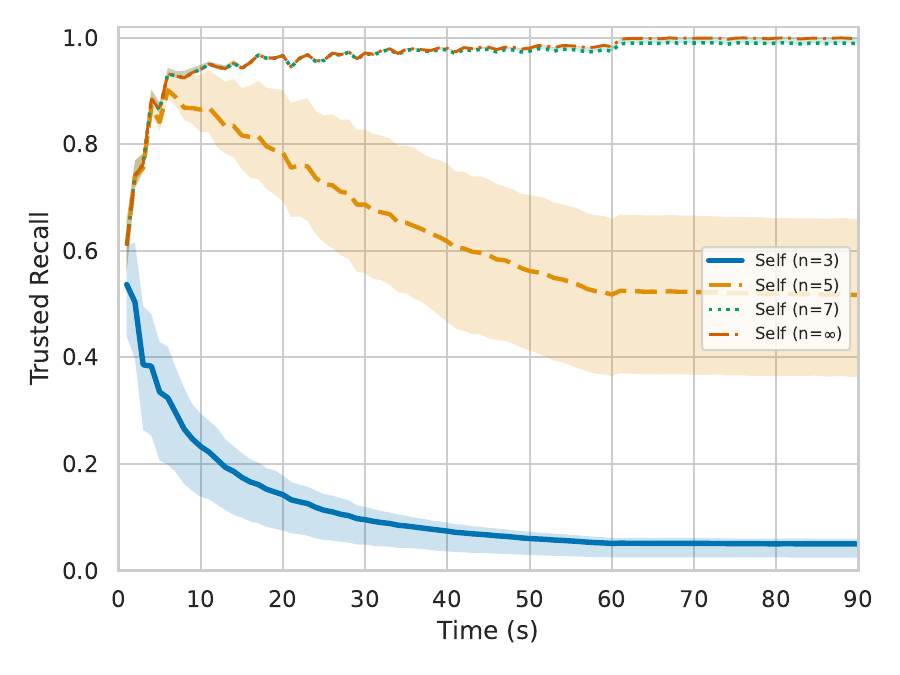}
        \caption{Self Anchor ($t=2$)}
        \label{fig:self_web2k_recall}
    \end{subfigure}
    \hfill
    \begin{subfigure}[t]{0.265\linewidth}
        \centering
        \includegraphics[width=\linewidth]{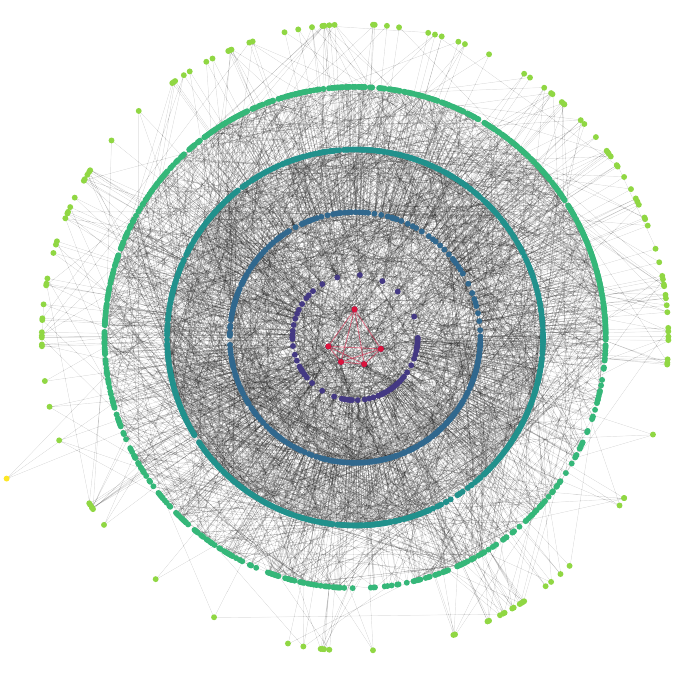}
        \caption{Certificate Graph ($t=2$)}
        \label{fig:web2k_graph}
    \end{subfigure}

    \caption{Average trusted recall in a 2000 agent collective where each agent requires two certificates to join with five genesis agents. Shaded regions show the IQR of trusted recall across agents.}
    \label{fig:web2k_all}
\end{figure}

Figure~\ref{fig:web2k_all} presents the same experiment with $t=2$ and a collective size of $2000$ agents. While trusted recall drops slightly compared to the $t=2$, $N=1000$ setting (Section~\ref{sec:eval_gov}), the absolute number of trusted agents increases, indicating that trusted sets grow with collective size. However, the drop in recall indicates that trusted set size grows sublinearly with collective size. 

\section{Related Work}

\paragraph{Scalable and decentralized multi-agent systems.} Recent work has explored architectures for connecting heterogeneous LLM agents beyond small, fixed teams. Internet of Agents (IoA) enables heterogeneous agents to discover and collaborate through an Internet-like architecture \citep{IoA}, while AgentNet and MACNET explore decentralized coordination and scalable multi-agent collaboration \citep{AgentNet,MACNET}. Symphony similarly targets scalable collective intelligence, although communication remains mediated by shared infrastructure \citep{symphony}. PPAI supports capability-aware agent selection and load balancing in a P2P population \citep{ppai}, while \cite{dgpan} study team formation and cooperation among agents. These systems primarily focus on discovery, routing, or a particular coordination mechanism. \panda instead solves for both of these and leverages its collective to improve fault tolerance. 

\paragraph{Multi-agent orchestration.}

A complementary line of work studies how agents should communicate once selected.
Centralized frameworks such as Magentic-One and MetaGPT impose predefined planning and coordination structures \citep{magentic-one,metagpt}, whereas DyLAN dynamically selects agents and communication structure for individual tasks \citep{dylan}. GPTSwarm represents agent workflows as optimizable computational
graphs \citep{gptswarm}, and Graph-of-Agents similarly performs agent selection and graph-structured message passing \citep{goa}. \panda separates the execution substrate from the orchestration strategy but draws inspiration from existing systems for its star, chain, and mesh topologies. G-Designer, AgentPrune, and AdaptOrch explore how to adaptively select an optimal MAS topology for a given task, these works could be integrated into \panda's protocol to allow for dynamic topology selection \citep{gdesigner, AgentPrune, adaptorch}. 

\paragraph{Fault tolerance and recoverable agent execution.}

Reliability in LLM-based MAS has largely been studied at the level of agent outputs and coordination errors. AgentScope provides robustness against faulty LLM or tool APIs \citep{AgentScope_2024}. \cite{whyMASfail} and \cite{ResilienceofMAS} show how MAS failures can arise from system design, inter-agent interaction, and verification failures, rather than solely from errors in individual model outputs. \cite{mas_bft_2026} and \cite{robustMAS} study resilience to faulty or Byzantine agents through consensus and ExoFlow explores execution and recovery for DAG-based workflows \citep{exoflow}. \panda additionally considers infrastructure failures and connects them with workflow recovery. 

\paragraph{Decentralized trust and governance.}
Governance for agentic systems commonly assumes a logically centralized provider. SAGA \citep{saga_ndss2026} enforces identity and interaction policies through a governance service, and subsequent work strengthens such governance against post-quantum and byzantine threats \citep{MAGIQ,saga_shield}. The NANDA index enables cryptographically verifiable discovery, but is designed as a one-time federated substrate to initiate interactions, not a collective architecture \citep{nanda}. \citet{garzon2025ai} anchor agent identities on a decentralized ledger, but still rely on credentials issued by commonly trusted third parties. Decentralized trust itself predates agents. SDSI proposed public-key-based authorization using certificates and linked local namespaces instead of a centralized hierarchy \citep{sdsi}, PGP \citep{pgp} introduced the concept of a web-of-trust and \cite{mas_trust} proposed distributed certificate-based trust for e-commerce applications. \panda builds on these ideas for LLM-based MAS by integrating decentralized trust into its collective.

\end{document}